%% file: main.tex
\documentclass[fleqn,10pt]{wlscirep}
\usepackage[utf8]{inputenc}
\usepackage[T1]{fontenc}

\input{preamble/preamble_math.tex}
\input{preamble/preamble.tex}
\input{preamble/preamble_acronym.tex}
\usepackage[skip=4pt plus1pt]{parskip}

\title{Robust Anomaly Detection for Particle Physics Using Multi-Background Representation Learning}

\author[1,+]{Abhijith Gandrakota}
\author[2,+]{Lily Zhang}
\author[2]{Aahlad Puli}
\author[3]{Kyle Cranmer}
\author[1]{Jennifer Ngadiuba}
\author[2]{Rajesh Ranganath}
\author[1]{Nhan Tran}
\affil[1]{Fermi National Accelerator Laboratory, Batavia, IL, USA 60510}
\affil[2]{New York University, New York, NY, USA, 10012}
\affil[3]{University of Wisconsin-Madison, Madison, WI, USA, 53706}

\begin{abstract}
Anomaly, or out-of-distribution, detection is a promising tool for aiding discoveries of new particles or processes in particle physics. In this work, we identify and address two overlooked opportunities to improve anomaly detection for high-energy physics. First, rather than train a generative model on the single most dominant background process, we build detection algorithms using representation learning from multiple background types, thus taking advantage of more information to improve estimation of what is relevant for detection. Second, we generalize decorrelation to the multi-background setting, thus directly enforcing a more complete definition of robustness for anomaly detection. We demonstrate the benefit of the proposed robust multi-background anomaly detection algorithms on a high-dimensional dataset of particle decays at the Large Hadron Collider. 
\end{abstract}
\begin{document}

\flushbottom
\maketitle
%
%
\thispagestyle{empty}

\section{Introduction}\label{sec1}

The Standard Model (SM) of particle physics has achieved remarkable success in predicting a wide range of phenomena, culminating in the recent groundbreaking discovery of the Higgs boson \cite{201230,chatrchyan_2013_observation,20121}. However, despite these triumphs, many phenomena in particle physics still remain unexplained \cite{Fukuda_1998, Muong-2:2023cdq, Barbier:2004ez}. Among the most important and exciting questions in physics is the discovery of phenomena beyond the Standard Model (BSM). This quest lies at the heart of the efforts at the Large Hadron Collider (LHC), dedicated to advancing humanity's understanding of the universe.

There are two main approaches for the discovery of new physics. The first, more traditional approach looks for a particular alternate hypothesis which is typically a family of parametrized distributions\cite{gluino2017}.
However, this approach seeks specific kinds of deviation and could be blind to other evidence of BSM physics not explicitly considered under the alternate hypothesis posed.

To address this shortcoming, a second, more recent approach uses machine learning-based anomaly detection (AD) algorithms to search for deviations from the standard model more broadly, without focusing on a specific BSM alternative hypothesis. 
There has been a growing interest in this data-driven approach in recent years, with the development of AD algorithms specific to high-energy physics~\cite{Park_2021, Dillon_2022, Canelli_2022, Hallin_2022, qcd_autoencoders, qcdorwhat} as well as their direct use in the analysis of data collected by the LHC experiements \cite{ATLAS:2020iwa,ATLAS:2023azi}. 


AD algorithms are employed to isolate collimated particle showers (jets) coming from the decays of new BSM particles produced by  proton proton collisions at the LHC. 
The hope is to discover a bump or a deviation in the smooth kinematic distributions such as the invariant mass of the selected jets (relative to what is otherwise expected from SM backgrounds), in order to provide evidence of a new BSM particle\cite{PhysRev.126.1858,20121,201230}. 
In other words, the discovery of a new particle can be viewed as hypothesis testing based on these kinematic distributions. 
To increase the power of this statistical test, the primary goal of the anomaly detection algorithm is to filter out SM backgrounds such that any deviations in these kinematic distributions resulting from a potentially new BSM process can be discerned. 
At the same time, this selection must be performed without creating deviations from the filtering process itself. 
This secondary objective is often known as decorrelation or the robustness of the anomaly detection algorithm \cite{Dolen_2016}. For further details please see \Cref{sec:background}.

In this work, we build upon existing data-driven approaches along two dimensions. The first is 
to take advantage of extra information and assumptions via representation learning fom multiple background processes. Existing approaches use deep generative models or density estimators such as variational autoencoders (VAEs) \cite{Kingma2013AutoEncodingVB,JimenezRezende2014StochasticBA} and normalizing flows (NF) \cite{Dinh2015NICENI, Dinh2017DensityEU} to model the most dominant SM process, quantum chromodynamics (QCD), as well as possible\cite{Workman:2022ynf}; however, AD methods based on generative modeling have been found to lead to worse than random chance failures in benchmark machine learning tasks \cite{nalisnick2018deep}, likely due to the sensitivity of such approaches to estimation error \cite{pmlr-v139-zhang21g}. One way to address this limitation of generative models and yield better anomaly detection is to transform the data inputs into representations that are easier to model and/or focus on information important for the task of discovery. In fact, in existing anomaly detection benchmarks on high-dimensional inputs, methods that build on the feature representations obtained from a model trained to distinguish non-anomalous data have been empirically shown to outperform detection based on generative models that directly estimate the non-anomalous input distribution \cite{salehi2021unified}. These empirical results do not guarantee that the former will always perform better than the latter, but they do point to the potential benefits of incorporating additional information or assumptions into anomaly detection. 
To incorporate additional information, we propose \textit{multi-background representation learning} for anomaly detection.
Our second contribution is developing a solution for robustness in the multi-background representation learning setting of anomaly detection. 
An important consideration for AD methods in jet physics is ensuring 
that anomaly scores do not depend on kinematic variables like the jet mass.
Otherwise, the probability of a false positive discovery increases. Concretely, discovery in high-energy physics often involves looking for statistically significant deviations or ``bumps'' in the overall sample of flagged anomalies along certain search variables, and such bumps are more likely when anomaly scores depend on these kinematic variables across jets within a known process.
To minimize such false positive discoveries, we introduce a decorrelation objective and algorithm in the presence of multiple backgrounds, taking inspiration from recent techniques in the robust representation learning and anomaly detection literature in machine learning \cite{puli2021out,zhang2023robustness}.

In this paper, we motivate and describe our robust multi-background representation learning approach to AD  and demonstrate empirically that it rivals the current state-of-the-art method in jet anomaly detection, specifically in detecting jets from top quarks as out-of-distribution samples given jets from QCD and W/Z processes as in-distribution examples.
We hope that this work inspires future work to consider underexplored ideas around multi-background representation learning for AD in high-energy physics.

\begin{figure}
    \centering
    \includegraphics[width=\linewidth,trim={0 120 0 80}]{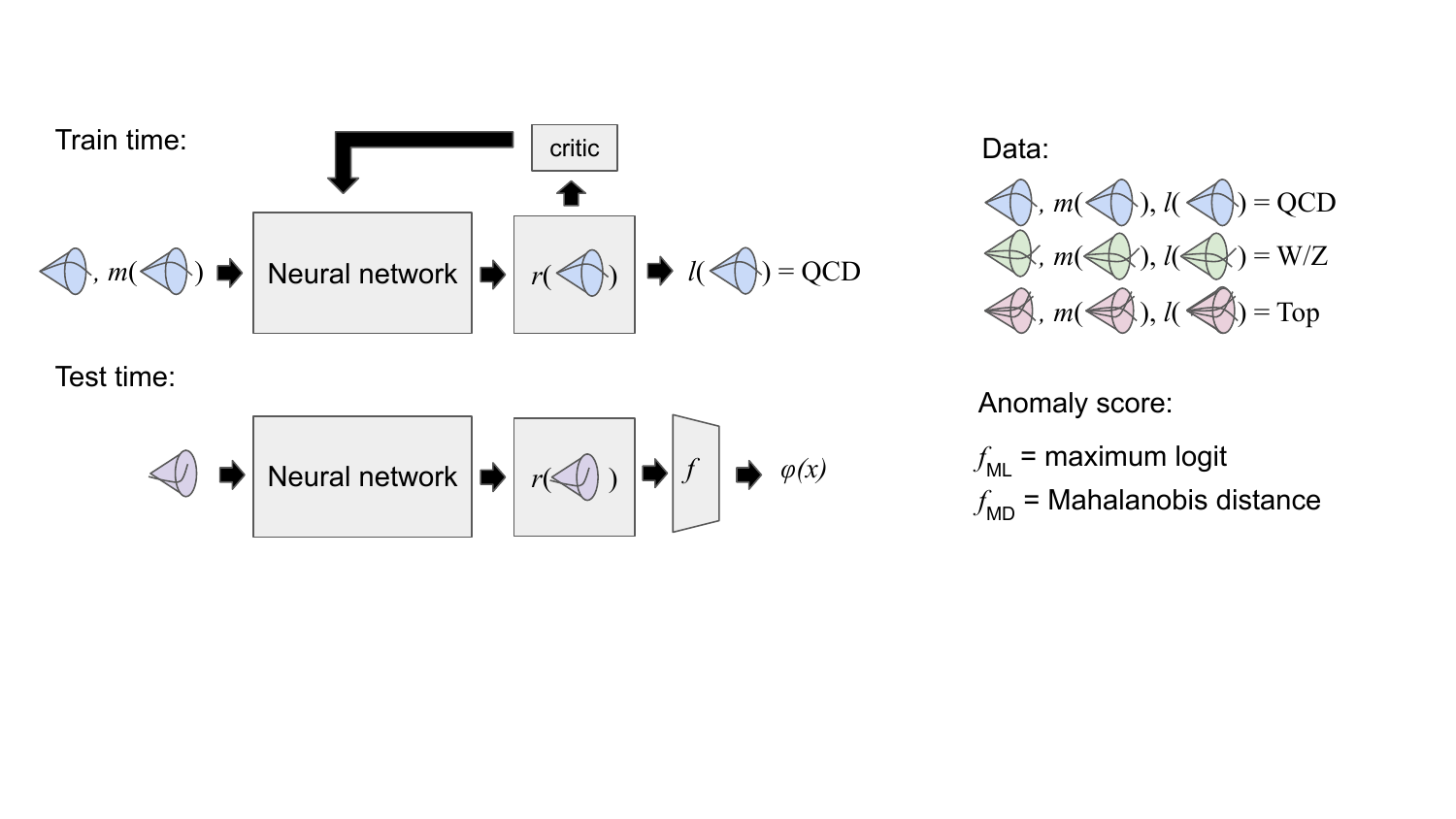}
    \caption{Overview of robust anomaly detection with multi-background representation learning. During training, jets (depicted by cones) as well as their mass $m$ and label $l$ are used to learn robust multi-background representations $r$. The data used for training includes jets of different background processes, not just QCD. Then, these learned representations are used to derive an anomaly score $\phi = f \circ r$ used at test time.}
    \label{fig:fig1}
\end{figure}
\section{Robust Multi-Background Anomaly Detection}\label{sec2}
Below, we motivate and describe our robust multi-background representation learning setup (\Cref{sec:rep_learn_mb}, \Cref{sec:rep_learn_na}) and overall anomaly detection algorithm based on such representations (\Cref{sec:anomaly_score}). A visual overview of the proposed method can be found in \Cref{fig:fig1}.

\subsection{Multi-Background Representation Learning}
\label{sec:rep_learn_mb}
The primary purpose of representation learning in anomaly detection is to guide the information used to assign an anomaly score. For instance, image anomaly detection algorithms built from representations that can distinguish known classes have been shown to outperform anomaly detection algorithms based on representations that do not take into account known class labels\cite{salehi2021unified}. In fact, among the latter, anomaly detection algorithms based on densities estimated from deep generative models have been shown to yield worse than random chance performance detecting anomalies that otherwise seem obvious based on human perception\cite{nalisnick2018deep, pmlr-v139-zhang21g}. These failures in some ways represent a worst-case scenario for purely data-driven approaches: without guidance on what types of deviations matter most, generative models 
can end up focusing modeling efforts on unimportant details at the expense of poorly estimating details that matter for detecting practical anomalies. This ability to learn from data while also incorporating  knowledge about what information is most important is where representation learning \cite{bengio2014representation} shines.
We build representations using multiple known background process types to better utilize the data already available from known physics. Denote the input as $\mbx$, the searchvariable we wish to be robust to (e.g., jet mass) as $\mbz$, and the background process as $\mby$. Define a representation function $r: \mathcal{X} \rightarrow \mathbb{R}^d$. 
A good representation retains relevant information for detection while removing information that simply yields an additional modeling burden to a downstream detection algorithm. We hypothesize that the  information that is relevant for detecting new particles will generally vary across known particles. To capture this assumption, we learn a representation that can distinguish between existing processes by training representations using the labels of the known particle classes as supervision.
In other words, we encourage representations that are useful for classifying between processes via $\arg\max_{r} p(\mby \g r(\mbx))$.

\subsection{Decorrelation under Multi-Background Representation Learning} 
\label{sec:rep_learn_na}
Second, we develop an approach for decorrelation under multi-background representations.
Namely, we directly enforce decorrelation within each background process by enforcing an independence constraint on the representations we learn described below. We use $\indep_p$ to denote independence of two random variables under distribution $p$. When there is no subscript specified, the independence is enforced under the original training distribution.
First, we introduce a distribution $\pind$ such that the search variable (e.g., jet mass) and label are independent: $\pind(\mbx, \mby, \mbz) = p(\mbx \g \mby, \mbz)p(\mby)p(\mbz)$. In other words, the distribution $\pind$ corresponds to the resulting distribution obtained after breaking dependence between $\mby$ and $\mbz$ occuring in the training distribution. 

Under this distribution, we first force representation $r(\mbx)$ to be independent of $\mbz$, i.e., $r(\mbx) \indep_\pind \mbz$. This constraint ensures that the representation cannot predict jet mass under a distribution where jet mass and label do not provide information about each other. It is important that this independence constraint be enforced under $\pind$, since enforcing marginal independence under the training distribution instead (i.e., $r(\mbx) \indep \mbz$) would disallow representations to contain information important for distinguishing particles, just because such information is also correlated with mass. 

Under $\pind$, we also enforce the representations to be independent of mass conditioned on the label, i.e., $r(\mbx) \indep_\pind \mbz \g \mby$.
This constraint ensures that jets belonging to a given background process cannot be differentially flagged as anomalies based on jet mass. Consequently, within a background process the distribution of jet mass will be the same for flagged jets as it is for all jets, an important condition for downstream analyses based on $\mbz$. For a more detailed discussion about downstream analyses and their assumptions, see \Cref{sec:background}.

Combining all the above objectives together, we have the following objective for the representations:
\begin{equation}\label{eq:desiderata}
\arg\max_{r} \pind(\mby \g r(\mbx)) \text{ s.t. }r(\mbx), \mby \indep_\pind \mbz.     
\end{equation}
The term $\arg\max_{r} \pind(\mby \g r(\mbx))$ encourages informative representations, while the constraint $r(\mbx), \mby \indep_\pind \mbz$ enforces decorrelation in the multi-background setting.
Note also that $r(\mbx), \mby \indep_\pind \mbz$ is the combination of $r(\mbx) \indep_\pind \mbz$ and $r(\mbx) \indep_\pind \mbz \g \mby$; moreover, for a given $r$, the predictor $\pind(\mby \g r(\mbx)) \text{ s.t. }r(\mbx), \mby \indep_\pind \mbz$ in \Cref{eq:desiderata} is equivalent to $p(\mby \g r(\mbx)) \text{ s.t. } r(\mbx) \indep \mbz | \mby$ (see \cite{zhang2023robustness} for proof).

To learn a representation $r$ that achieves \Cref{eq:desiderata}, we use Nuisance\footnote{In prior work\cite{puli2021out,zhang2023robustness}, $\mbz$ was called a nuisance variable, not to be confused with nuisance parameter which is much more common in the high-energy physics context. To avoid confusion, in this work we refer to $\mbz$ as a search feature, given its role in the discovery process.} Randomized Distillation (NuRD) \cite{puli2021out}. First, to approximate the distribution $\pind$ from the training distribution, we perform importance weighting: 
\begin{equation}\label{eq:rw}
 \pind(\mbx, \mby, \mbz) = w_{\mby,\mbz}p(\mbx, \mby, \mbz), \quad w_{\mby,\mbz} = p(\mby)/p(\mby | \mbz).
\end{equation}
Both $p(\mby)$ and $p(\mby \g \mbz)$ can be estimated from the training data, assuming access to both the jet mass and label. 

To encourage joint independence, we use a mutual information penalty. Namely, we estimate $\mathbf{I}_\pind(r(\mbx), \mby; \mbz)$, where any nonzero quantity suggests the lack of joint independence.
To estimate this mutual information, we employ the density ratio trick \cite{Sugiyama2012DensityRE}, following Ref.~\cite{puli2021out}. Concretely, we train an additional neural network, referred to as the critic model, to estimate the relevant density ratio for mutual information. The critic model $p_\gamma$ is trained to distinguish between inputs from $\pind(r(\mbx),\mby,\mbz)$ and inputs from $\pind(r(\mbx), \mby)\pind(\mbz)$, the latter of which we represent by shuffling the jet masses within the batch. The critic model predicts unshuffled inputs with $c=1$ and shuffled inputs with $c=0$. Then, given well-calibrated critic model such that the probability $p_\gamma(c = 1 \g r(\mbx),\mby,\mbz)$ corresponds to the probability of the input coming from distribution $\pind(r(\mbx),\mby,\mbz)$, we can estimate the mutual information as follows: 
\begin{equation}\label{eq:mi}
\mathbf{I}_\pind(r(\mbx), \mby; \mbz)  =  
\mathbb{E}_{\pind(r(\mbx), \mby, \mbz)} \big[ \log p_\gamma(c = 1 \g r(\mbx),\mby,\mbz) - \log( 1-p_\gamma(c = 1 \g r(\mbx),\mby,\mbz)) \big].
\end{equation}
To approximate the mutual information under $\pind$, we weight the examples using the weights in \Cref{eq:rw}. 

\begin{figure}
    \centering
    \includegraphics[width=0.9\textwidth]{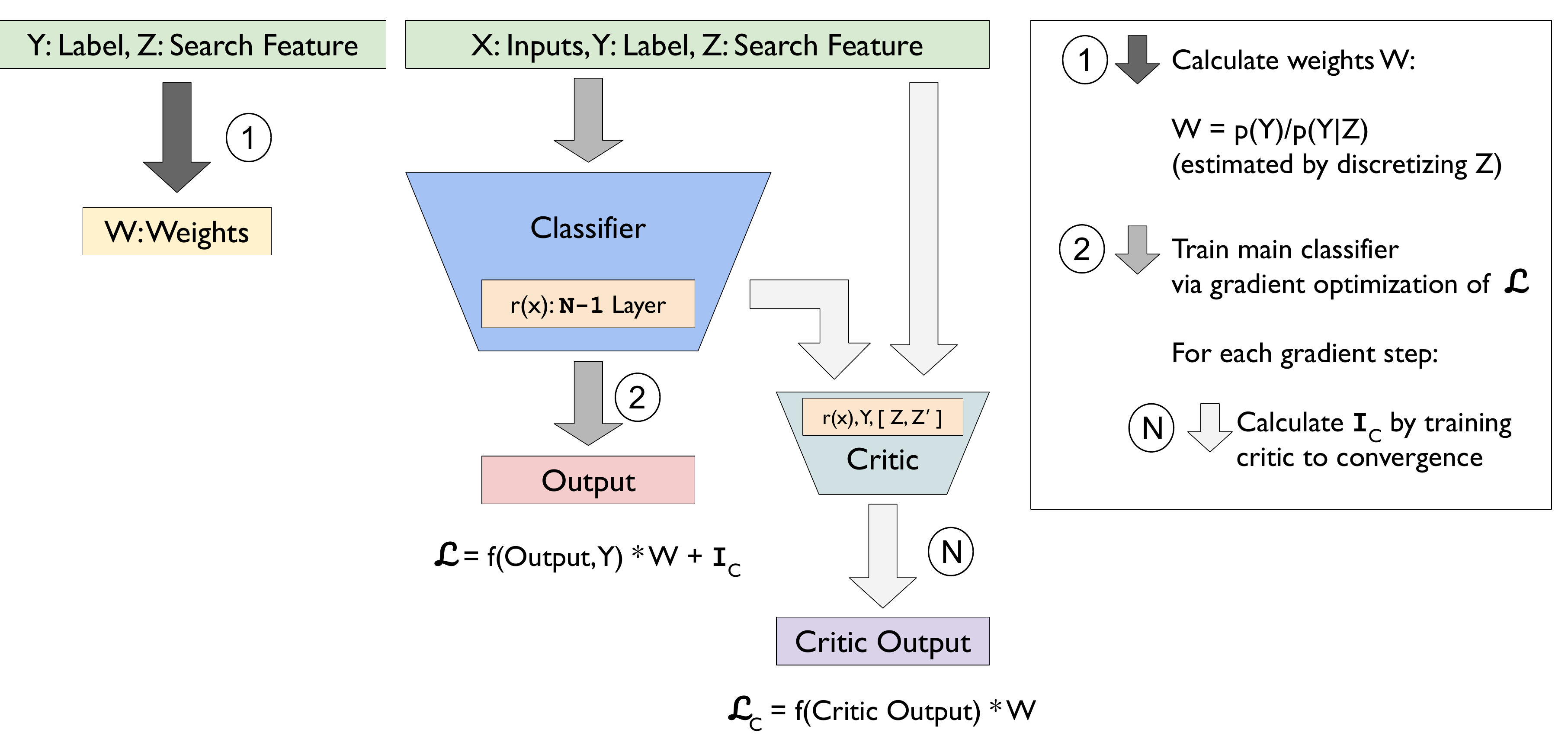}
    \caption{An overview of the Nuisance-Randomized Distillation algorithm \cite{puli2021out} used in this work for learning robust multi-background representations for anomaly detection in high-energy physics. First, we calculate weights $w$ for each input based on its label $y$ and jet mass $z$. These weights are used to approximate the distribution $\pind$ such that the $y$ and $z$ are marginally independent. Then, we train our classifier to optimize a reweighted objective. We additionally include a mutual information-based penalty term to the loss by training a critic model $n$ gradient steps for every gradient step of the classifier.}
    \label{fig:method}
\end{figure}

Putting both ideas together,
we learn a representation $r$
by optimizing parameters $\theta$ via gradient descent under the following objective:
\begin{equation}\label{eq:nurd}
    \max_\theta \quad \log \pind(\mby \g r_\theta(\mbx)) - \lambda \mathbf{I}_\pind(r_\theta(\mbx), \mby; \mbz).
\end{equation}
In other words, our  objective seeks a representation that distinguishes between known background process types as well as possible while penalizing representations that together with the background process label can predict the search feature. 
See \Cref{fig:method} for a schematic of the overall algorithm.


\subsection{Anomaly Scores based on Robust Multi-Background Representations}
\label{sec:anomaly_score}
Given a representation built using multiple backgrounds (i.e., more than one distinct value of $\mby$) and the appropriate decorrelation of mass with respect to all of them, 
we define a robust multi-background detection score $\phi: \mathcal{X} \rightarrow \mathbb{R}$ as a function of such a representation, i.e., $\phi(\mbx) = f(r(\mbx)), \; f: \mathbb{R}^d \rightarrow \mathbb{R}$. Here, we describe two anomaly scores, $\phi_{ML}$ and $\phi_{MD}$, which utilize the aforementioned robust multi-background representations. In particular, we define $\phi_{ML} = f_{ML} \circ r$ and $\phi_{MD} = f_{MD} \circ r$, where functions $f_{ML}$ and $f_{MD}$ map the representation to some scalar anomaly score. 
Following the deep learning anomaly detection literature, we define two anomaly scores yielding high accuracy on image anomaly detection tasks; the first, the max logit (ML) \cite{hendrycks2022scaling}, takes the maximum logit of the classifier trained on in-distribution inputs and labels. The second, the Mahalanobis distance (MD) \cite{lee2018simple}, fits class-conditional Gaussians to the data distribution representations and uses the probability under this fitted distribution as the anomaly score. We describe both in more detail below.

The max logit anomaly score flags a jet as an anomaly if the maximum of the logits is small \cite{hendrycks2022scaling}. Since a good classifier on the data distribution of known classes should generally assign high maximum logits to jets from known classes, a jet that is assigned relatively small outputs relative to jets seen during training is likely different in some way (e.g. does not activate the same feature maps as strongly) and thus more likely to be anomalous. To implement the max logit score, we use the classifier that trains the representation described in the previous section.

The Mahalanobis distance focuses instead on the representations of individual jets.  
Namely, the Mahalanobis distance models the feature representations of in-distribution data as $k$ class-conditional Gaussians with means $\mu_k$ and covariances $\Sigma_k$. At test time, the anomaly score is the minimum Mahalanobis distance from a new input's feature representations to each of these class distributions, 
$\min_k \sqrt{(r(\mbx) - \mu_k) \Sigma_k^{-1}(r(\mbx) - \mu_k)^\top} = \max_k p(r(\mbx)\mid\mby=k)$. 
Assuming 
minimal overlap in probability across each class-conditional Gaussian (e.g., representations for each class form tight clusters that are far apart), this method 
can approximate density estimation on the representations: $\max_k p(r(\mbx) \mid \mby=k) \propto \max_k p(r(\mbx)\mid\mby=k)p(\mby=k) \approx \sum_k p(r(\mbx)\mid\mby=k)p(\mby=k) = p(r(\mbx))$. 

While density estimation based on Gaussians for each known class may seem relatively simple, this method is especially powerful coupled with a good representation that clusters similar jets. Moreover, the speed of inference, especially relative to running a more complex deep generative model, makes this score especially practical given the computational burden of processing petabytes of collision data~\cite{Duarte:2018ite}.

We call the above methods \gls{nurd-ml} and \gls{nurd-md}. These two anomaly detection methods based on robust representations have been shown to offer significant performance gains in out-of-distribution benchmarks relative to approaches without these robustness considerations\cite{zhang2023robustness}.

Summarizing the full method, 
first we learn representations via the NuRD objective (\Cref{eq:nurd}) by training a classifier to predict the background particle classes under a reweighted objective (\Cref{eq:rw}) with a mutual information penalty (\Cref{eq:mi}), implemented with an inner submodule that estimates a mutual information term. Then, we derive two anomaly scores from these representations, one which involves fitting class-conditional Gaussians on the representations, and the other which takes the maximum logit score from the classifier derived from the representations. See \Cref{fig:fig1} for an overview of the entire detection algorithm pipeline.

\section{Related Work}\label{sec:related}
Our multi-background and representation learning-based approach is related to and inspired by work in the anomaly and out-of-distribution detection literature in high-energy physics and machine learning. The move away from purely generative-based approaches towards an approach that considers other techniques is motivated by existing work on the limitations of deep generative models for anomaly detection \cite{nalisnick2018deep,pmlr-v139-zhang21g}. The use of class labels of in-distribution processes to improve anomaly detection is inspired by existing work in the image out-of-distribution detection literature which shows that the most performant methods take advantage of label information \cite{Salehi_2021}. In particle physics applications, existing works have considered the idea of building classifiers for representation learning~\cite{Cheng:2022gma}, as well as embedding data into lower-dimensional spaces while preserving different choices of metrics\cite{Park:2022zov}.

The need for robust anomaly detection in high-energy physics has been discussed in \cite{qcdorwhat, Louppe2016}. Several approaches have been proposed for encouraging detection that is decorrelated with kinematic variables such as jet mass \cite{Heimel2018QCDOW,Dolen_2016,disco}. 
Ref.~\cite{Heimel2018QCDOW} trains an autoencoder with reconstruction loss and an additional adversarial term which penalizes the loss if an adversary (a neural network) can predict the jet mass from the prediction residual. In this approach, the objective encourages reconstruction error to be independent of the jet mass, which is equivalent to enforcing the independence between the anomaly score and the jet mass. Given that only QCD is considered, this can also be trivially viewed as a form of conditional independence, i.e. $\phi(\mbx) \indep \mbz | \mby$.
Another approach to ensure robustness is post-hoc decorrelation, which chooses different thresholds for classification depending on jet mass. A common approach is to bin the data based on jet mass and then choose thresholds per bin such that the false positive rate is the same for all bins \cite{Dolen:2016kst}. This objective is the same as the equal opportunity constraint (i.e., $\phi(\mbx) \indep \mbz | \mby = 1$) in the fairness in machine learning literature, which is a relaxation of the full conditional independence constraint $\phi(\mbx) \indep \mbz | \mby$, also known as equalized odds in the fairness literature \cite{hardt2016equality}. 
Practically, a post-processing method can be applied on top of any detection method, including the one we propose. Moreover, our predictor also satisfies conditional independence, which means that it generalizes the objectives of both aforementioned approaches. 

There also exist various methods which incorporate decorrelation into the classification of jets. Ref.~\cite{Louppe2016} employs an additional adversarial term to encourage the distribution of a classifier's predictions to be invariant to a nuisance parameter either conditional on the class $\mby$ or with class marginalized out.
Rather than using an adversarial objective, Ref.~\cite{disco} incorporates a distance correlation penalty which is zero if and only if the relevant variables are independent.
The goals of jet classification are distinct from those of anomaly detection, but this work demonstrates how some of these ideas can be ported over to anomaly detection when the latter makes use of supervised representation learning.

\section{Experimental Setup}\label{sec:experiments}
To demonstrate the practical benefit of our method, we train and test \gls{nurd-ml} and \gls{nurd-md} on simulated high-momentum jets at the LHC. We first describe the data used to benchmark our approach (\Cref{sec:data}). Then, we describe details of the model architecture and training (\Cref{sec:models}). We compare our results with a baseline VAE approach as defined in \cite{Heimel2018QCDOW}, described in related work in \Cref{sec:related}. To test the benefits of the approach, we keep architectural choices between our proposed approach and baseline as close as possible; namely, we design our classifier to match the architecture of the VAE encoder so that empirical benefits cannot be attributed to architectural choices.

\subsection{Data}\label{sec:data}
The model is trained on high momentum jets from the hls4ml LHC Jet dataset set given in Refs.~\cite{Moreno_2020, pierini_maurizio_2020_3602260}. The jets are represented as an image of $50 \times 50$ pixels, where each pixel corresponds to the sum of the energy of all particles in the corresponding cell. These images are normalized into a probability distribution over the grid (i.e. pixel values sum to one), as in this task we are uninterested in the total energy within the grid but rather the spatial distribution over the grid. We train a classifier on two different standard model jet classes, QCD and W/Z bosons. For testing the algorithm, we use jets produced by top quarks as the out-of-distribution sample. 
For constructing the in-distribution dataset, we used 600,000 jets with equal proportions of QCD and W/Z boson jets. We split this dataset into training, validation, and test sets with proportions 60\%, 20\%, and 20\% respectively. We use the training and validation data to train the classifier to obtain a representation function for anomaly detection.
For evaluating anomaly detection performance, we use a dataset consisting of the test QCD jets alongside out-of-distribution top jets. 


\subsection{Model Architecture and Training}\label{sec:models}

Our main classifier takes as input the $50\times 50$-size images as input and predicts the particle class (QCD or W/Z) as output. This classifier is a convolutional neural network whose architecture closely follows that of the encoder in Ref.~\cite{qcdorwhat} for a fair comparison of methods. We train our classifier on a dataset comprised of QCD and W/Z jets in equal proportions, in order to bypass the optimization difficulties of a highly imbalanced distribution that training on a dataset of natural proportions would introduce. 

We train the NuRD algorithm as described in Section~\ref{sec2} which modifies traditional supervised learning by additionally reweighting the examples to approximate $\pind$ and estimating the mutual information term $\text{I}_\pind(r(\mbx), \mby; \mbz)$ to enforce joint independence. 

To reweight the examples, we discretized the jet mass into bins of 5 GeV for both the QCD and W/Z samples to accommodate the resolution of the W/Z resonance peaks. Then, the weight given to a jet in the bin $i$ is the inverse of the probability of that bin under the distribution of the jet mass for that class: more specifically, we use weights $N/n_i$, where $N$ is the total number of jets in the class and $n_i$ number of jets in bin $i$. This is equivalent to using the weight $\frac{p(\mby)}{p(\mby \g \mbz)}$ described in the \Cref{sec2}, except that the jet mass has been discretized. We utilize these weights to estimate the empirical risk under $\pind$ when computing the loss over a batch.

To estimate the mutual information penalty, we initialize a critic model which takes in the 20-dimensional representation $r(\mbx)$ from the penultimate activations of the main classifier, as well as a scalar label and jet mass. The network is a multilayer perceptron (MLP) with 3 hidden layers of 256, 128, and 64 neurons and predicts which distribution the input came from (i.e.,  $\pind(r(\mbx),\mby,\mbz)$ or $\pind(r(\mbx), \mby)\pind(\mbz)$).

For each step of the main classifier, we train and update the critic on a 10\% fraction of the dataset and use the resulting critic model to approximate the mutual information per batch. This term is added as a penalty to the reweighted cross-entropy loss, which we optimize via gradient descent using the Adam optimizer with learning rate of $10^{-3}$.

We train a baseline VAE on 300,000 QCD jets, using the network architecture and the hyperparameters as defined in Ref.~\cite{Heimel2018QCDOW}. We evaluate both proposed methods \gls{nurd-ml} and \gls{nurd-md}, and the baseline VAE method on a test set containing only QCD and out-of-distribution top-quark jets, following previous evaluations \cite{Heimel2018QCDOW}.

\section{Results}\label{sec:results}

\begin{figure}[htbp]
    \centering
    \subfigure[ROC Curve]{\includegraphics[width=0.33\textwidth]{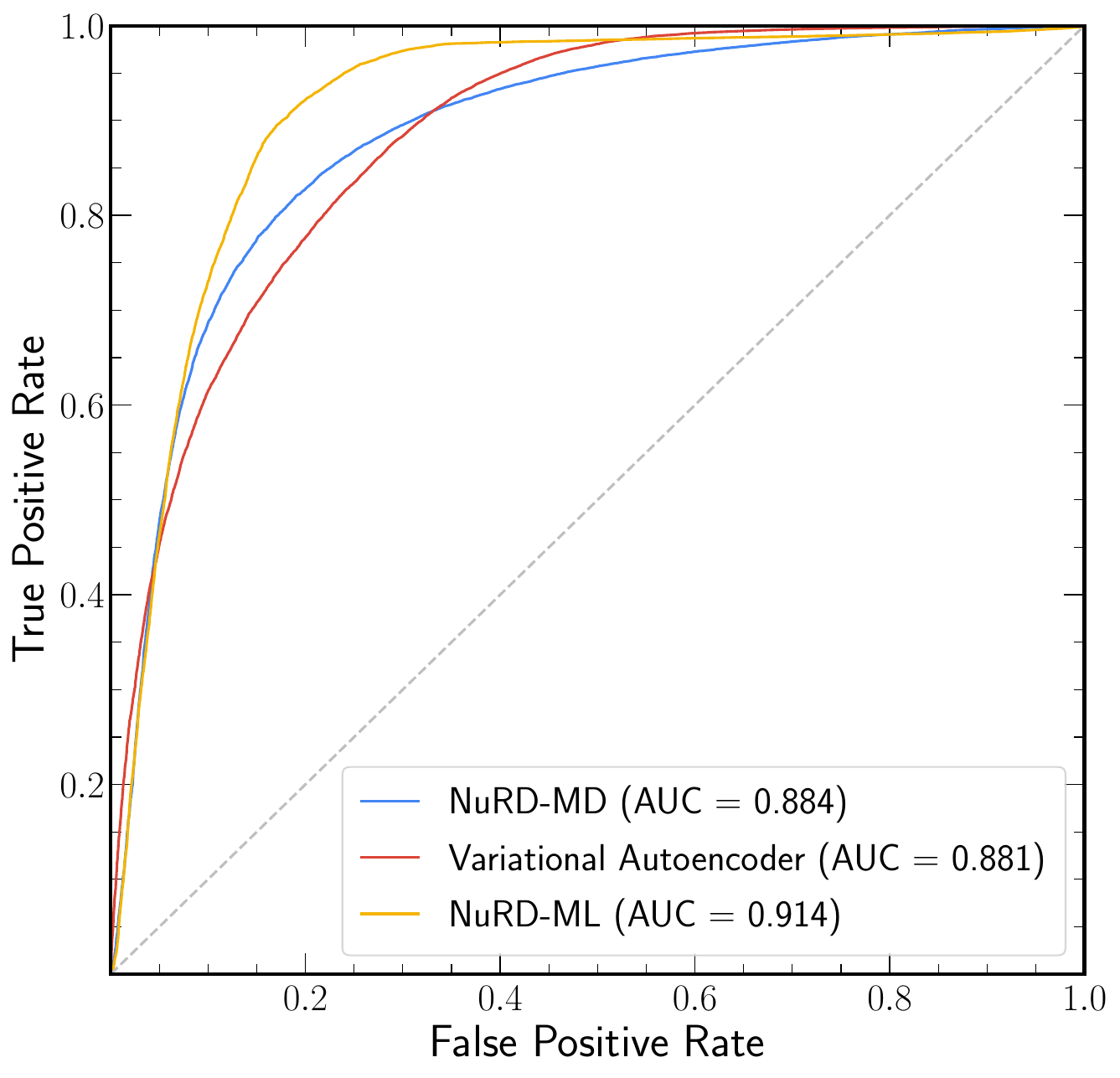}}
    \subfigure[Mass distribution]{\includegraphics[width=0.33\textwidth]{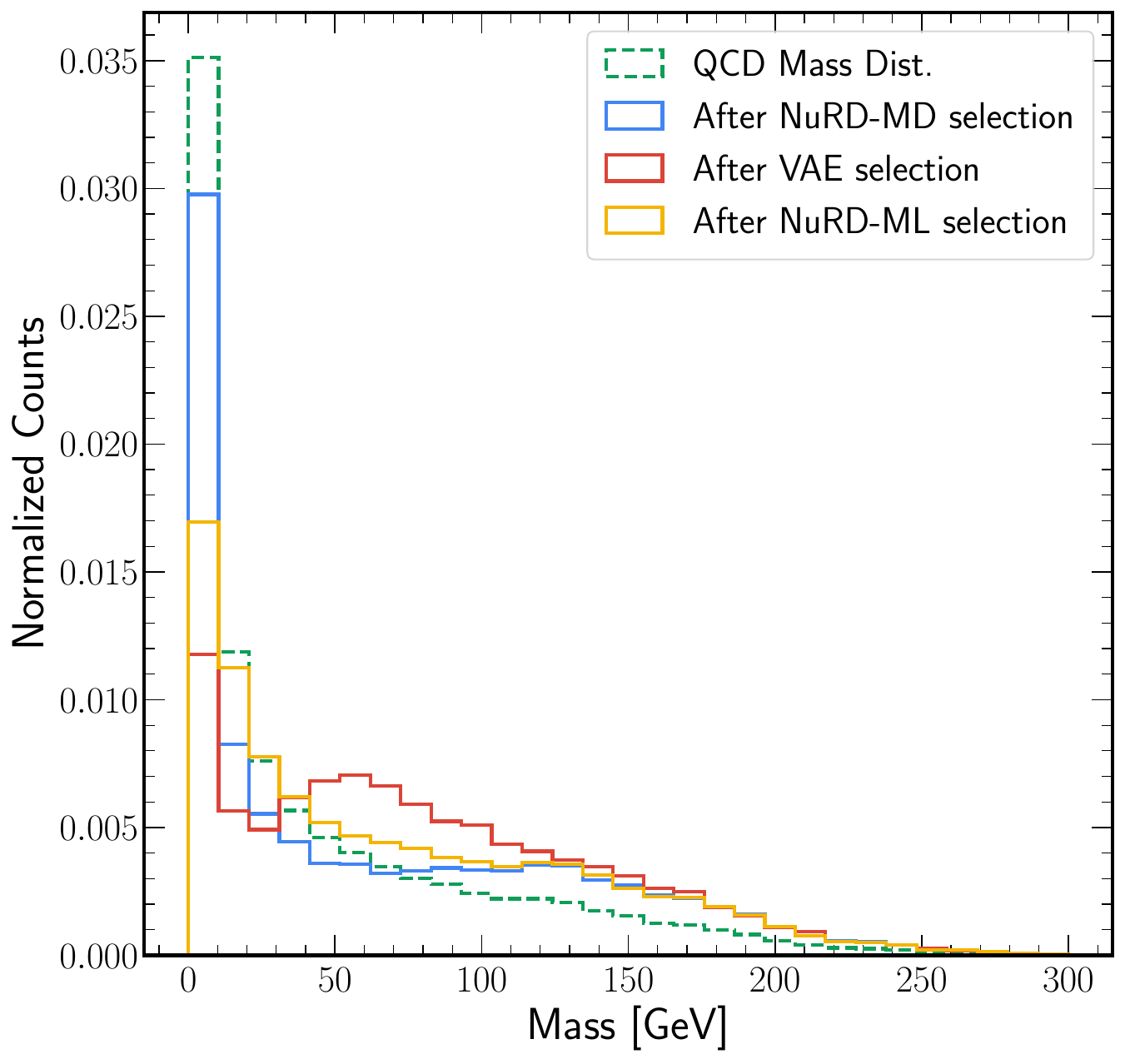}}
    \caption{Our proposed robust multi-background detection methods outperform the baseline VAE implementation of Ref.~\cite{qcdorwhat} in both the overall detection performance (AUROC, left) as well as decorrelation with jet mass (less sculpting, right).}
    \label{fig:main}
\end{figure}

\Cref{fig:main}(a) shows that \gls{nurd-ml} and \gls{nurd-md} yield better detection performance than the baseline VAE method as measured via AUROC. 
Recall that both the baseline and \gls{nurd-md} perform anomaly detection via density estimation, one directly on the inputs and the other on a representation of the inputs learned from multiple types of background processes. Given this parallelism, the superiority of \gls{nurd-md} despite its use of a simpler generative model (i.e. class-conditional Gaussians rather than a deep generative model) suggests that a well-crafted representation can assist in anomaly detection by making the task of density estimation easier. The fact that \gls{nurd-ml} also outperforms the baseline suggests that the learned representation can be useful even for anomaly scores that do not directly correspond to density estimation.

Moreover, \Cref{fig:main}(b) shows the QCD mass distribution overall (green) and after subsetted to the 50\% most anomalous events. 
An ideal algorithm would show no difference in this mass distribution before and after thresholding, such that downstream analyses could treat the flagged anomalies in aggregate using known physics about QCD. On the other hand, a poor algorithm results in a ``sculpting'' of the mass distribution into one unlike the original QCD mass distribution, increasing the potential for false positive discoveries.
Both \gls{nurd-md} and \gls{nurd-ml} yield less sculpting of the jet mass distribution than the baseline VAE method. The degree of sculpting is quantified in \Cref{tab:main} which reports the Jensen-Shannon divergence and L2 Wasserstein Distance between the jet mass distributions before and after applying the threshold on the anomaly score. 
The fact that the proposed methods achieve better detection performance as well as less sculpting of the QCD mass distribution suggests they are a superior option to the baseline. The presence of some amount of sculpting from the proposed methods is likely due to the use of a penalty in the objective rather than a hard constraint.

\Gls{nurd-ml} and \gls{nurd-md} also exhibit an overall better significance improvement (\Cref{tab:main}).
The significance improvement for a given true positive rate is defined as the signal efficiency (true positive rate) divided by the square root of the background misidentification (false positive rate). This metric is indicative of the discovery potential~\cite{pdg}.
\Cref{{tab:main}} shows the maximum significance improvement versus the signal efficiency. \gls{nurd-ml} and \gls{nurd-md} yield a better signal-to-background ratio, indicating they falsely flag a smaller proportion of background jets for a given proportion of anomalous jets flagged. Both proposed methods also have a higher maximum significance improvement, suggesting they can achieve a better optimal signal-to-background ratio with a high signal efficiency. 

\begin{table}[htbp]
    \centering
    \begin{tabular}{lcccc}
    \toprule
    Method  & 
    AUROC ($\uparrow$)  & 
    JSD ($\downarrow$) & 
    L2 WD ($\downarrow$) & 
    SI ($\uparrow$) \\
    \hline

   VAE & 0.881 & 0.255 & 34.3 & 2.03\\
   \hline
   \gls{nurd-ml} & \underline{\textbf{0.914}} & \textbf{0.168} & \textbf{24.4} & \underline{\textbf{2.32}} \\
    \gls{nurd-md} & \textbf{0.884} & \underline{\textbf{0.118}} & \underline{\textbf{19.1}} &  \textbf{2.23} \\
    \bottomrule
    \end{tabular}
    \caption{Main results comparing the baseline with proposed methods. We look at the area under the ROC curve (AUROC) when classifying in-distribution from out-of-distribution samples; the Jensen-Shannon Divergence (JSD) and L2 Wasserstein distance (L2 WD) between the QCD mass distribution before and after 
    filtering to the 50\% most anomalous inputs;
    and the maximum significance improvement (SI). We bold all results better than baseline and underline the best performance.
    }
    \label{tab:main}
\end{table}

\begin{figure}[htbp]
    \centering
    \subfigure[Representations under first two principal components.]{\includegraphics[width=0.33\textwidth,trim={0 1cm 0 0},clip ]{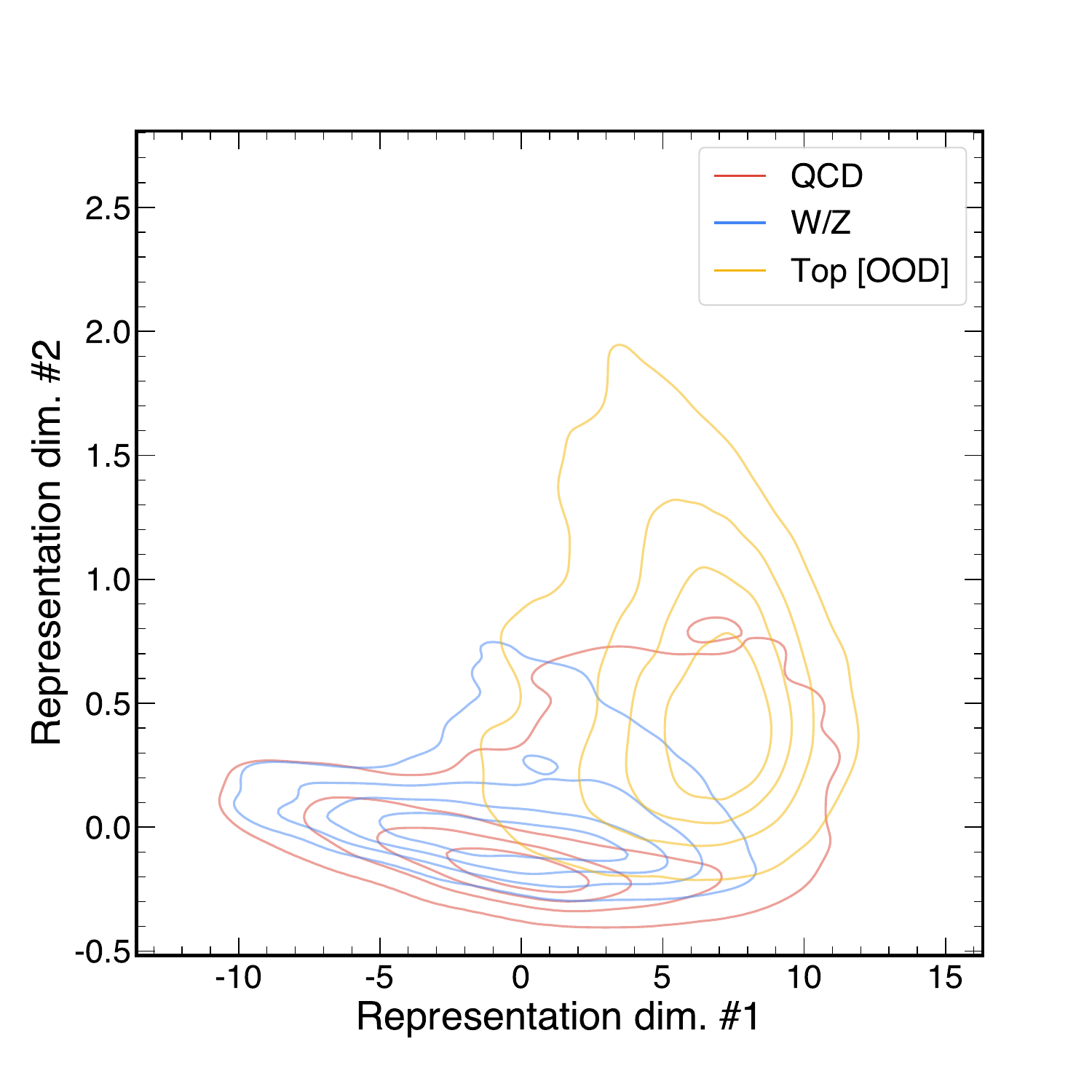}}
    \subfigure[The Mahalanobis distance from the representations.]{\includegraphics[width=0.33\textwidth,trim={0 1cm 0 0},clip]{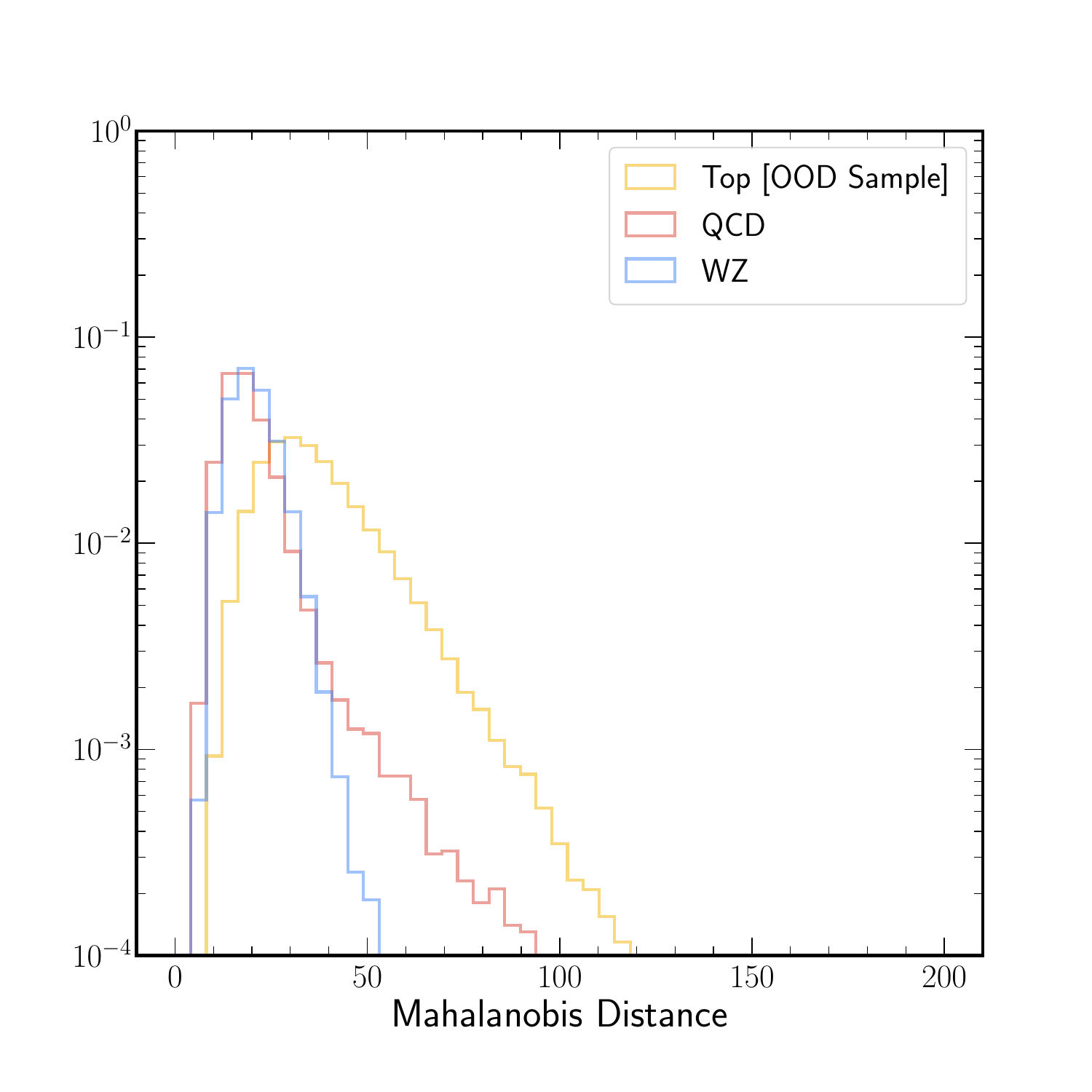}}
    \subfigure[Max Logit score for backgrounds and OOD]{\includegraphics[width=0.28\textwidth]{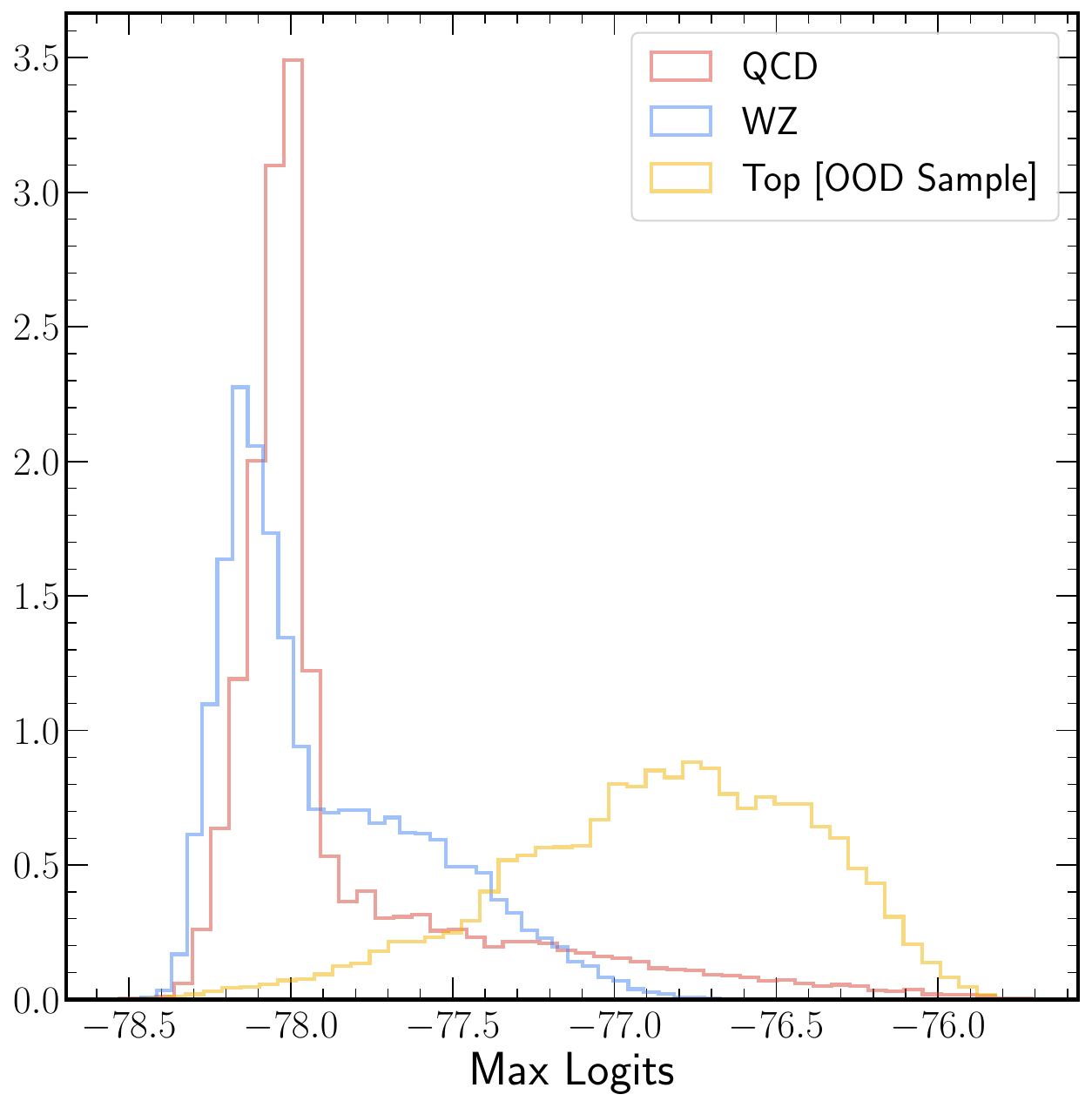}}
    \caption{(a) PCA on the learned representations show that they yield good separation of the out-of-distribution top process and the in-distribution QCD and W/Z processes. This leads to good separation of the downstream anomaly detection scores, Mahalanobis distance (b) and Max Logit (c). 
    }
   \label{fig:pca_rep}
\end{figure}

Finally, we visually assess the quality of our learned representation. We perform a Principle Component Analysis (PCA) on the 20-dimensional representation and visualize the results of the first two principal components in \Cref{fig:pca_rep}(a). We observe that the representations of the out-of-distribution top-quark jets are well-separated from those of the in-distribution QCD and W/Z jets.
The Mahalanobis distance computed from these representations 
also show a clear separation of OOD samples from the backgrounds (\Cref{fig:pca_rep}(b)), as does the Max Logit scores (\Cref{fig:pca_rep}(c)).


\section{Summary}

We present a new approach for anomaly detection in high energy physics via representation learning from multiple background distributions. Our approach takes advantage of more data than existing data-driven approaches by considering multiple jet types as well as classification labels distinguishing them. In addition, we take inspiration and motivation from the out-of-distribution detection and spurious correlations literature in machine learning to develop and motivate decorrelation in the multi-background setting. Our experiments illustrate empirically that the proposed robust multi-background representation learning approach yields better detection and decorrelation than state-of-the-art VAE-based detection\cite{Heimel2018QCDOW} while keeping architectural decisions fixed. Based on these results, we encourage future work in anomaly detection for high-energy physics to consider multi-background approaches to improve the potential for discovery of new science.
\clearpage
\section*{Acknowledgements}

AG, JN, and NT are supported by Fermi Research Alliance, LLC under Contract No. DE-AC02-07CH11359 with the Department of Energy (DOE), Office of Science, Office of High Energy Physics. 
JN is also supported by the U.S. Department of Energy (DOE), Office of Science, Office of High Energy Physics ``Designing efficient edge AI with physics phenomena'' Project (DE-FOA-0002705). JN, NT are also supported by the DOE Office of Science, Office of Advanced Scientific Computing Research under the ``Real-time Data Reduction Codesign at the Extreme Edge for Science'' Project (DE-FOA-0002501). This work was supported in part by the AI2050 program at Schmidt Futures (Grant G-23-64934). NT is also supported by the DOE Early Career
Research program under Award No. DE-0000247070.
LZ, AP, and RR are partly supported by NSF
Award 1922658 NRT-HDR: FUTURE Foundations, Translation, and Responsibility for Data Science, NSF CAREER Award 2145542, ONR N00014-23-1-2634, and Google. AP is also partly
supported by the Apple Scholars in AI/ML PhD fellowship.

\section*{Author contributions statement}

\bibliography{biblio}

\appendix
\section*{Appendix}
\section{Background \& Motivation}
\label{sec:background}
Here, we provide introductory background on anomaly detection in high-energy physics for non-physics readers. We first describe the common approach of bump hunting (\Cref{sec:bumphunt}). Then we describe how existing approaches handle knowledge of multiple known processes (\Cref{sec:multiple}). Finally, we motivate our proposed approach within the introductory framework described below (\Cref{sec:connection}).
\subsection{Bump hunting} 
\label{sec:bumphunt}
The existing paradigm of machine learning-based particle discovery typically proceeds as a bump hunt, which can be formalized as a hypothesis test based on the aggregate collection of jets in a given collision and a statistic such as jet mass. Concretely, the test asks whether the mass distribution of jets for a given experiment looks like the expected mass distribution of known background processes $P_0$, or whether one should reject this null hypothesis in favor of an alternative that additionally considers the presence of a new particle that creates a bump in the otherwise expected mass distribution~\cite{CMS:2022usq}. This hypothesis test considers jets in aggregate because individual jets alone do not provide sufficient evidence for new physics in the presence of noise from external factors. The test focuses on mass rather than some other combination of statistics so that it can provide actionable information for theories priors which themselves consider mass (e.g., as a key property of a new particle).

So where does an anomaly detection algorithm operating on a per-jet basis come into play? First, considering all jets in aggregate makes it difficult to discern mass deviations from a novel particle that is only present in very small proportions; consequently, it is useful to first filter out as many jets from known processes as possible to allow jets from a potential new particle to make up a larger proportion of the overall distribution. This filtering process is performed via  an anomaly detection algorithm: given an anomaly score $\phi: \mathcal{X} \rightarrow \mathbb{R}$, a detection score $d: \mathcal{X} \rightarrow \mathbb{R}$ is generally of the form $d(\mbx) = \mathbf{1}[\phi(\mbx) > \tau]$. The better the algorithm is at filtering out known processes (low false positive rate) while flagging novel processes to keep around in the hypothesis testing phase (high true positive rate), the more power the resulting hypothesis test has to find a deviation and thus a new particle. To maximize the accuracy of the detection algorithm, generally the high-dimensional information describing a jet is incorporated into this algorithm.

A key requirement for the anomaly detection algorithm is that the expected mass distribution under the null hypothesis must still be known for the jets remaining after the filtering step. To meet this requirement, one option is ensure that the distribution of mass does not change after filtering: i.e., $P_0(\mbz|d(\mbx) = 1)= P_0(\mbz)$.  One way to enforce this constraint is to define mass-based thresholds $\tau_{\mbz}$ for filtering. Alternatively, one can also force the upstream anomaly detection scoring function to be independent of mass for known jets: $\phi(\mbx) \indep_{P_0} \mbz$, implying $d(\mbx) \indep_{P_0} \mbz$ and thus $P_0(\mbz|d(\mbx) = 1)= P_0(\mbz)$.

In summary, we have the following procedure describing existing bump hunts. Let $(\mbx, \mby, \mbz)$ denote the high-dimensional information, process label, and mass of a jet. Then, we have the following hypothesis test, where $d(\mbx) \indep_{P_0} \mbz$:
\begin{align*}
    &H_0: \{ \mbz | d(\mbx) = 1\} \sim P_0(\mbz) \\
    &H_A: \{ \mbz | d(\mbx) = 1\} \not\sim P_0(\mbz).
\end{align*}
\subsection{How existing approaches handle multiple known SM processes}
\label{sec:multiple}
Thus far we have abstracted out the details of this known reference distribution $P_0$. This distribution is in fact a mixture distribution of several processes, including QCD, W/Z, and Top quark processes; considering just these three processes, we have $P_0 = a P_{\text{QCD}} + b P_{\text{W/Z}} + c P_{\text{Top}}$ with known mixture coefficients. Generally, existing methods build anomaly detection algorithms on QCD jets only. Then, one of two procedures follows: either $P_{\text{QCD}}$ is used as an approximation of $P_0$ in the above procedure, since $a >> b >> c$, or the other mixture components are accounted for in the hypothesis testing phase with a few assumptions, which we discuss next.

There exist very accurate approximations of $P_{\text{W/Z}}(\mbz), P_{\text{Top}}(\mbz)$, and the distribution of kinematic variables in general for processes other than QCD. Given $P_{\text{W/Z}}(\mbz), P_{\text{Top}}(\mbz)$ are known and $P_{\text{QCD}}(\mbz)$ can be estimated from simulated data, we can estimate the filtered mass distribution $P_0(\mbz|d(\mbx) = 1)$ assuming the detection algorithm satisfies decorrelation within each process, i.e., $d(\mbx) \indep \mbz | \mby$ for each $\mby \in \mathcal{Y}_{\text{known}}$:

\begin{align}
    P_0(\mbz|d(\mbx) = 1) &= \sum_{\mby \in \mathcal{Y}_{\text{known}}} P(\mbz, d(\mbx) = 1, \mby) / P(d(\mbx = 1)) \\
    &=\sum_{\mby \in \mathcal{Y}_{\text{known}}} P(\mbz|d(\mbx) = 1, \mby)P(d(\mbx) = 1|\mby)P(\mby) / P(d(\mbx = 1)) \\
    &=\sum_{\mby \in \mathcal{Y}_{\text{known}}} P(\mbz|\mby)P(d(\mbx) = 1|\mby)P(\mby) / P(d(\mbx = 1)) \label{eq:conditional_indep}\\
    &=\frac{a P_\text{QCD}(\mbz)P(d(\mbx) = 1|\mby=\text{QCD})
        + b P_\text{W/Z}(\mbz)P(d(\mbx) = 1|\mby=\text{W/Z})
        + c P_\text{Top}(\mbz)P(d(\mbx) = 1|\mby=\text{Top})}{\int P(\mbz, d(\mbx = 1))d\mbz}. \\
    &=\frac{a P_\text{QCD}(\mbz)P(d(\mbx) = 1|\mby=\text{QCD})
        + b P_\text{W/Z}(\mbz)P(d(\mbx) = 1|\mby=\text{W/Z})
        + c P_\text{Top}(\mbz)P(d(\mbx) = 1|\mby=\text{Top})}{ aP(d(\mbx) = 1|\mby=\text{QCD})+ bP(d(\mbx) = 1|\mby=\text{W/Z}) + cP(d(\mbx) = 1|\mby=\text{Top})}.\label{eq:final}
\end{align}
In other words, one can accurately estimate the mass distribution of flagged jets under the null by knowing 1. the distribution of mass for each process, 2. the relative proportions of each process for the overall experiment, and 3. the detection algorithm's false positive rates for each known process type, assuming decorrelation is met. 
Note that all of 1, 2, 3 are either known through theory / accurate simulation or can be estimated, e.g. $P(d(\mbx)=1|\mby)$ can be estimated by evaluating the detection algorithm on training data. Then, a key lever for improving the probability of successful scientific discovery is reducing the false positive rate of flagging known processes as anomalies (while successfully flagging true anomalies).
Additionally, while existing works assume $d(\mbx) \indep \mbz | \mby$ for each $\mby \in \mathcal{Y}_{\text{known}}$, they do not enforce it directly since the detection algorithms are trained on QCD jets only.


\subsection{An alternative approach to increase power and reduce false positive discovery}
\label{sec:connection}
As with any test, the goal of the above process is to have a test with high power and a low false discovery rate (FDR). The former goal of high power is achieved when the anomaly detection algorithm has a high true positive rate of flagging anomalies and a low false positive rate of flagging existing processes as anomalous. The result is then that novel jets make up a larger proportion of the flagged jets, making it easier for any test to detect their presence (in the form of a deviation from the expected mass distribution of known processes). The latter goal of low FDR is the purpose of decorrelation, to ensure that any deviations in the mass distribution are a result of new unknown processes rather systematic biases in the detection algorithm which "sculpt" the mass distribution away from its expected shape. In this work, we introduce robust multi-background anomaly detection and show how it can help both increase power and reduce false positive discoveries. Below, we provide a brief summary of the source of the gains.

First, rather than training a detection algorithm on QCD jets alone and assuming that other jets are consistently flagged as non-QCD, multi-background anomaly detection algorithms learn from data from all known jets to improve the rate at which all are filtered out. 
Representation learning guides what information is most important to learn, and in this work, we focus on the information that distinguishes known jets under the hypothesis that this information is most important to distinguish existing jets from new jets as well. If this is true, then such multi-background representation learning can increase overall power of the hypothesis test of interest.

Next, rather than assuming decorrelation is achieved across all jet types, we directly enforce decorrelation within each known background process, i.e. $d(\mbx) \indep \mbz | \mby$ for each $\mby \in \mathcal{Y}_{\text{known}}$. Doing so ensures that the distribution of a kinematic variable such as jet mass is the same for a given background process whether we are considering flagged jets or all jets, i.e. $P(\mathbf{z} | \mathbf{y}, d(\mathbf{x})) = P(\mathbf{z} | \mathbf{y})$. This equivalence in the distribution of jet mass makes it possible to filter down the mass distribution of flagged jets further based on knowledge of $P(\mathbf{z} | \mathbf{y})$ obtained from accurate simulations. To achieve this decorrelation, we enforce an independence constraint on the representations we learn. Then, detection algorithms based on these representations are less susceptible to false discoveries than algorithms that do not enforce that decorrelation for all background processes.

\end{document}

%% file: preamble/preamble_math.tex
\newcommand{\mbx}{\textbf{x}}
\newcommand{\mby}{\textbf{y}}
\newcommand{\mbz}{\textbf{z}}

\newcommand{\g}{\,\vert\,}

\newcommand{\indep}{\perp \!\!\! \perp}

\newcommand{\pind}{{p_{\scaleto{\indep}{4pt}}}}

%% file: preamble/preamble.tex
\usepackage{soul}
\usepackage{hyperref}
\usepackage{cleveref}
\usepackage[acronym,shortcuts,nowarn,nohypertypes={acronym,notation}]{glossaries}
\usepackage{scalerel}
\usepackage{subfigure}

%% file: preamble/preamble_acronym.tex
\newacronym{nurd-ml}{nurd-ml}{Nuisance Randomized Distillation with Max Logit}
\newacronym{nurd-md}{nurd-md}{Nuisance Randomized Distillation with Mahalanobis Distance}

\newacronym{ood}{ood}{out-of-distribution}